\documentclass[aps,floatfix,twocolumn,showpacs]{revtex4}
\usepackage{graphicx,amssymb,amsmath}
\begin{document}
\title{Entanglement-Secured Single-Qubit Quantum Secret Sharing}
\author{P. Scherpelz}
\altaffiliation[Current address:  ]
{Department of Physics, University of Chicago, 5720 S. Ellis Ave, Chicago, IL 60637.}
\author{R. Resch}
\altaffiliation[Current address:  ]
{SLAC National Accelerator Laboratory, 2575 Sand Hill Road, Menlo Park, CA 94025-7015.}
\author{D. Berryrieser}
\altaffiliation[Current address:  ]
{Department of Applied Physics, 348 Via Pueblo Mall, Stanford University Stanford, CA 94305-4090.}
\author{T. W. Lynn}
\email{lynn@hmc.edu}
\affiliation{Department of Physics, Harvey Mudd College, 301 Platt Blvd., Claremont, CA, 91711.}
\date{\today}

\begin{abstract}

In single-qubit quantum secret sharing, a secret is shared between $N$ parties via manipulation and measurement of one qubit at a time.  Each qubit is sent to all $N$ parties in sequence; the secret is encoded in the first participant's preparation of the qubit state and the subsequent participants' choices of state rotation or measurement basis.  We present a protocol for single-qubit quantum secret sharing using polarization entanglement of photon pairs produced in type-I spontaneous parametric downconversion.  We investigate the protocol's security against eavesdropping attack under common experimental conditions:  a lossy channel for photon transmission, and imperfect preparation of the initial qubit state.  A protocol which exploits \textit{entanglement} between photons, rather than simply polarization \textit{correlation}, is more robustly secure.  We implement the entanglement-based secret-sharing protocol with 87\% secret-sharing fidelity, limited by the purity of the entangled state produced by our present apparatus.  We demonstrate a photon-number splitting eavesdropping attack, which achieves no success against the entanglement-based protocol while showing the predicted rate of success against a correlation-based protocol.
\end{abstract}
\pacs{03.67.Hk, 03.67.Dd}
\maketitle

\section{Introduction}
Secret sharing is the general term for a communication task in which one participant (the sender) wants to share a message with multiple other participants (the recipients) in a way that forces the recipients to cooperate with one another to reconstruct the message.  The task is relevant when recipients are considered more trustworthy as a group than individually.  In the strongest version of secret sharing, the message can be fully reconstructed by the full set of $N-1$ recipients; however, any subset of $N-2$ or fewer recipients possesses zero information regarding the message.  This task can be implemented classically by distributing between $N-1$ recipients $N-2$ randomly generated bit strings, or `shadows,' and a final $(N-1)$th `shadow' string which is the bitwise sum or XOR of the other $N-2$ strings and the original message.  All $N-1$ recipients together can reconstruct the message by taking the XOR of their shadows; however, any proper subset of recipients possesses no information about the message.

Classical secret sharing protocols generally do not involve the secure transmission of the shadows, leaving this task to cryptographic protocols.  The emerging field of quantum cryptography hinges on the secure transmission of information encoded in quantum states.  Thus, by using quantum states to encode and distribute the shadows, secure communication can be built into a secret sharing protocol; quantum-state resources enhance the sharing of classical bit-string sequences.  Such quantum-mechanical resources can be brought to the task of secret sharing in several distinct ways.

The original quantum secret sharing protocol, presented in 1999 \cite{qss_orig}, requires the use of a multipartite entangled state.  Specifically, to share a single-bit secret between a sender and $N-1$ recipients, the $N$-qubit entangled state $\frac{1}{\sqrt{2}}[|0\rangle_1|0\rangle_2\ldots|0\rangle_{N} + |1\rangle_1|1\rangle_2\ldots|1\rangle_{N}]$ must be produced, and the individual qubits distributed between the participants.  This quantum secret sharing protocol is in principle quite powerful.  It allows participants to share secrets composed not only of bit values (0 or 1) but of complete qubits (quantum states of the form $a|0\rangle + e^{i\varphi}b|1\rangle$).  The production of multipartite entangled states is unfortunately a technical challenge, requiring an experimental tour de force at each realization \cite{multipartite1,multipartite2,multipartite3,multipartite4, multipartite5}.

Single-qubit quantum secret sharing (SQQSS), by contrast, was first proposed and demonstrated in 2005 \cite{Schmid, Schmid2} using photon pairs produced by type-II spontaneous parametric downconversion (SPDC).  The protocol involves the transmission of a single qubit $a|0\rangle + e^{i\varphi}b|1\rangle$ through the entire sequence of participants.  The sender prepares a state with a specific value of $\varphi$ and each recipient performs a simple operation to alter $\varphi$.  The final participant performs a measurement on the qubit whose outcome depends on the final value of $\varphi$, and the secret -- the initial $\varphi$ value -- can be reconstructed only when all recipients reveal their individual operations.  Such a protocol uses quantum resources to allow secure sharing of a classical secret, but does not enable the more powerful sharing of a full quantum-state secret.  On the other hand, SQQSS relies on physical states which can be easily produced and manipulated in the laboratory, allowing for the straightforward realization of the protocols and demonstration of their successes and vulnerabilities.  The original SQQSS protocol, with proposed precautions and coding repetitions \cite{Schmid_corr,HeWang}, provides security against numerous cheating attacks by a subset of recipients. 

In this work, we experimentally implement two variations on the protocol of \cite{Schmid, Schmid2}, adapted for use with type-I SPDC.  One version, like the original, relies only on polarization correlation in photon pairs; the other directly exploits the quantum entanglement between photons.  We note the relative strengths and weaknesses of the two versions.  In particular, we develop and implement a photon-number splitting (PNS) eavesdropping attack; in the presence of a lossy transmission channel and imperfect state preparation, this eavesdropping attack works against the correlation-based protocol but fails against the entanglement-secured version.

\section{Single-Qubit Quantum Secret Sharing Schemes Using Type-I SPDC}

\subsection{Correlation-Based Protocol}
\label{subsec:Correlation-Based Protocol}

The SQQSS protocol relies on the secure transmission of one qubit to a number of participants sequentially.  In order to prevent individual participants from cheating, however, this \textit{signal} qubit must be produced and detected in correlation with a partner qubit, called the \textit{idler}.  In both the original version \cite{Schmid} and our own variation, the qubits are encoded in the polarizations of a pair of photons produced in SPDC.  Thus henceforth we will refer specifically to polarization states of photons rather than to generic two-state quantum systems.  Our correlation-based protocol for type-I SPDC closely follows the original treatment of \cite{Schmid} for the type-II case.

The sharing of a secret begins with the creation of a pair of photons in the polarization-entangled state
\begin{equation}
	|\psi_0\rangle=\frac{1}{\sqrt{2}}(|HH\rangle+|VV\rangle)
	\label{eq:spdcIstate}  
\end{equation}
where $H$ denotes horizontal polarization and $V$ vertical polarization of each photon.  The idler photon passes through a polarizer oriented to transmit $|H\rangle$, while the signal photon passes through a chain of SQQSS participants.  Taken on its own, the polarization state of the signal photon as it enters the SQQSS chain is undefined.  However, the final step in the protocol is the detection of the signal and idler photons in coincidence with one another.  This coincidence detection projects the signal photon, at its entry into the SQQSS chain, into the state $|H\rangle$.  

\begin{figure}[t]
	\begin{center}
		\includegraphics[width=3.125in, trim = 0in 0in 0in 0in, clip=true]{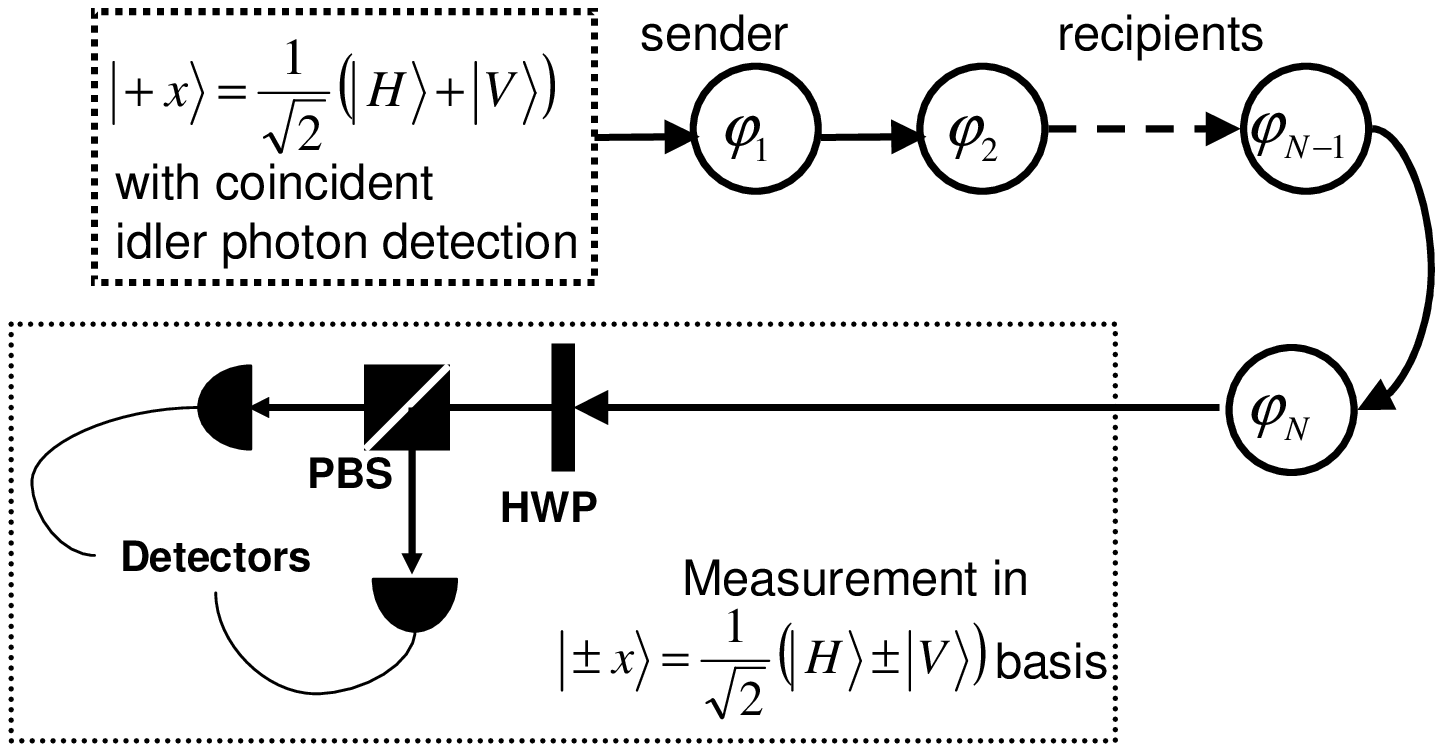}
		\caption{A schematic of the single qubit quantum secret sharing protocol. The qubit is initially in the state $|+x\rangle=\frac{1}{\sqrt{2}}(|H\rangle+|V\rangle)$, conditioned on coincident detection of the idler photon (gating).  Each participant applies a relative phase shift $\varphi_j \in \{0,\pi/2,\pi,3\pi/2\}$ to the $|V\rangle$ component. The half-wave plate (HWP) and polarizing beam splitter (PBS) allow for measurement of the final state in the $|\pm x\rangle = \frac{1}{\sqrt{2}}(|H\rangle \pm |V\rangle)$ basis.
		\label{fig:overallschem}}
	\end{center}
\end{figure}

The SQQSS chain is shown conceptually in Fig. \ref{fig:overallschem}.  It begins with the sender, who uses a combination of waveplates to transform the signal photon to one of the four states
\begin{align}
|+x\rangle &= \frac{1}{\sqrt{2}}(|H\rangle+|V\rangle), \notag \\
|+y\rangle &= \frac{1}{\sqrt{2}}(|H\rangle+i|V\rangle), \notag \\
|-x\rangle &= \frac{1}{\sqrt{2}}(|H\rangle-|V\rangle), \\
|-x\rangle &= \frac{1}{\sqrt{2}}(|H\rangle-i|V\rangle) \notag .
	\label{eq:xydefined}
\end{align}
This is equivalent to the preparation of $|+x\rangle$ followed by use of a tilt-adjustable phase plate to obtain one of ${|\pm x\rangle,|\pm y\rangle}$, as shown in Fig. \ref{fig:overallschem}.
The signal photon then passes to $N-1$ recipients in turn; each one uses a tilt-adjustable phase plate to apply a randomly selected phase shift $\varphi_j \in \{0,\pi/2,\pi,3\pi/2\}$ so that after participant $k$, the signal photon state is
	\begin{equation}
	|\chi_k\rangle = \frac{1}{\sqrt{2}}(|H\rangle+\mathrm{exp}(i\displaystyle\sum_{j=1}^{k}{\varphi_{j}})|V\rangle).
	\label{eq:statek}
	\end{equation}
Overall, $j$ ranges from $1$ to $N$ to account for the sender ($i=1$) and $N-1$ recipients.  The set of four possible phase shifts can be broken down into two classes:
\begin{align}
\mathrm{Class} \, X &\Rightarrow \varphi_j \in \{0,\pi\} \notag \\
\mathrm{Class} \, Y &\Rightarrow \varphi_j \in \{\pi/2,3\pi/2\}.	  
\label{eq:classes}
\end{align}
The sender's choice of state preparation likewise falls into either Class $X$, for creation of $|\pm x\rangle$, or Class $Y$, for creation of $|\pm y\rangle$.

Thus each participant so far possesses a single `class' bit denoting which class they have chosen, but also a second `secret' bit consisting of their phase choice within that class.  After $N$ participants have applied their local operations, the initial signal photon state $|H\rangle$ is transformed to the state
\begin{equation}
	|\chi_{N}\rangle = \frac{1}{\sqrt{2}}(|H\rangle+\mathrm{exp}(i\displaystyle\sum_{j=1}^N \varphi_j)|V\rangle).
	\label{eq:endstate}
	\end{equation}
These phase changes are illustrated in Fig. \ref{fig:overallschem}.

At this point, the final participant measures the polarization of the signal photon in the $|\pm x\rangle = (1/\sqrt{2})(|H\rangle\pm|V\rangle)$ basis, conducting the measurement in coincidence with the idler photon as specified earlier.  The measurer records the measurement outcome, and the physical aspect of the SQQSS is complete.

Notice that if the final state is of the form
	\begin{equation}
	|\chi_N\rangle = \frac{1}{\sqrt{2}}(|H\rangle\pm|V\rangle),
	\label{eq:finalstatex}
	\end{equation}
then the measurement outcome will be $|+ x\rangle$ or $|-x\rangle$, each with probability 1.  However, if
\begin{equation}
	|\chi_N\rangle = \frac{1}{\sqrt{2}}(|H\rangle\pm i|V\rangle),
	\label{eq:finalstatey}
	\end{equation}
	the final measurement result will be random.
	
Therefore, after the measurement is performed, the $N$ participants publicly announce the class of the operation they applied, and the total number of Class $Y$ operations is counted.  If the number of Class $Y$ operations was even, the run is valid; the $N-1$ recipients can expect to reconstruct the secret from their shadows.  If the number of Class $Y$ operations was odd, then the run is discarded; this happens half the time on average.
	
For valid runs, the remaining `secret' bit value retained by each participant regarding their applied operation constitutes that participant's shadow (or the sender's secret).  The last participant also holds the record of the measurement outcome.  Only if the $N-1$ recipients share their shadows can they determine the sender's secret, $\varphi_1$.

Finally, to prevent cheating, a random subset of the bits must be checked.  To do this, the first $N$ participants announce in random order their actual phase changes $\varphi_j$ for a subset of runs randomly selected by the sender.  The expected measurement result for each run is computed and compared to the measurement result announced by the final participant.  If any recipient attempts to cheat by measuring the single-photon state and sending a newly prepared version along to subsequent participants, the bit error rate rises to at least 25\%.  Thus, if the protocol shows an error rate of less than 25\%, cheating of this form can be ruled out.  

Restrictions on the order of the class announcements, along with repetitions of the protocol with coding enhancements, can be implemented to make the protocol more secure, defending even against cheaters with their own entangled-pair resources \cite{HeWang}.  Alternately, the protocol can defend against cheating by a subset of recipients (participants 2 through $N$) via a simpler modification:  instead of measuring the signal photon, participant $N$ is required to transmit it back to the sender, participant 1.  The sender then measures the photon in either the $|\pm x\rangle$ or $|\pm y\rangle$ basis, but all recipients announce their class choices before the sender announces the sending and measurement classes.  Runs with even numbers of Class $X$ operations, \textit{including the measurement class}, are considered valid.  This protocol defends against cheating, even with entangled-pair resources, by any subset of the recipients.  It privileges the trusted sender of the message, but the sender already occupies a position of trust by knowing the original secret, in many if not all possible applications.

\subsection{Entanglement-Based Protocol}
\label{subsec:Entanglement-Based Protocol}

In both type-I protocols, the signal and idler photons are first generated as the entangled pair of Eq. \ref{eq:spdcIstate}.  In the correlation-based protocol, all measurements are done in coincidence with detection of the idler photon in the state $|H\rangle$.  Thus the signal photon is projected into the state $|H\rangle$ as well, and then rotated into $|\pm x\rangle$ or $|\pm y\rangle$ state afterwards by means of phase-shifting optics.  As pointed out in Refs. \cite{Schmid, Schmid2}, the polarization correlation between the signal and idler photons is crucial for the security of the protocol; a cheater or eavesdropper, who does not have access to the idler photon, therefore has no information on the initial polarization of the signal photon.  Because the signal photon's initial state is ill-defined, subsequent measurements by the cheater cannot reveal information about the phase changes applied by sender or recipients.  

However, while the correlation-based protocol relies on the polarization correlation between signal and idler, it does not rely explicitly on the quantum entanglement between them.  This insensitivity to entanglement \textit{per se} can be viewed as a strength of the protocol, giving robustness against imperfect entanglement in the form of a lack of coherence between the two terms in the superposition of Eq. \ref{eq:spdcIstate}.  However, by failing to fully exploit the quantum entanglement in the initial resource the correlation-based protocol passes up a chance for enhanced security against eavesdropping attacks, as we demonstrate in the next section.

The entanglement-based protocol which follows makes full and explicit use of the entanglement between signal and idler in order to prepare the signal photon in the state $|+x\rangle$ before it enters the chain of SQQSS participants.  Unlike a classical mixture, the initial entangled state can be rewritten as
\begin{equation}	|\psi_0\rangle=\frac{1}{\sqrt{2}}(|HH\rangle+|VV\rangle)=\frac{1}{\sqrt{2}}(|+x,+x\rangle+|-x,-x\rangle).
	\label{eq:twobases}  
\end{equation}
Relying on this equality, we orient the idler polarizer to transmit only $|+x\rangle$ polarization.  Detection of the signal photon in coincidence with the idler then projects the signal photon into the $|+x\rangle$ state as it enters the SQQSS chain. The sender and recipient roles remain the same as before, except that the measurement of the signal photon is now conducted in coincidence with the $|+x\rangle$-selected idler.

The success of the entanglement-based protocol depends entirely on the presence of entanglement, rather than classical correlation, between the signal and idler polarizations.  To the extent that the initial state is a classical mixture of $|HH\rangle$ and $|VV\rangle$, projection of the idler photon onto $|+x\rangle$ will leave the signal photon in an uncertain polarization state rather than projecting it onto $|+x\rangle$.  Thus the success rate, or fidelity, of the secret-sharing transmission is sensitive to the purity of entanglement in this protocol.  However, the explicit use of entanglement makes this protocol robust against certain eavesdropping attacks and cheating strategies to which the correlation-based protocol is vulnerable.  We present one such attack, a photon-number splitting exploitation of experimental asymmetries, in the next section.

\section{Photon-Number Splitting Eavesdropping Attack}
\label{sec:PNS Eavesdropping Attack}

The correlation-based SQQSS protocol, in an ideal realization, provides security against eavesdropping.  However, the protocol remains vulnerable to attacks that may arise due to the imperfections of implementation.  In quantum communication in general (for example, in quantum key distribution), these forms of attack have been some of the greatest obstacles to securely implementing protocols \cite{qcrypt,practical_qcrypt1,practical_qcrypt2}.  Such attacks include the possibility of splitting off and measuring a fraction of the photons in a pulse that is sent for each qubit (photon-number splitting), or adding photons to the channel and later extracting them for measurement (Trojan-Horse attacks) \cite{trojan}.  Here we focus on robustness against a photon number-splitting, or PNS, attack.

In a PNS attack, an eavesdropper takes advantage of the fact that implementations may involve transmission of pulses with the possibility of multiple photons per qubit.  The eavesdropper `splits' off some of the photons, and measures the state of their polarization, while the remaining photons pass through to the intended receiver untouched.  Unless the intended receiver can detect the decreased signal size, the eavesdropper gains information about the message without alerting the participants to her presence.

There are many ways of avoiding PNS attacks in quantum key distribution, including decoy states, strong reference pulses, and differential phase shifts \cite{decoys1,decoys2}.  The correlation-based protocol protects against PNS by referencing the signal photon to the idler photon, which is passed through a polarizer before being detected.  A PNS eavesdropper cannot discover whether her 'picked-off' signal photons are coincident with the correct idler photons; thus she gains no information about the polarization state of the signal \cite{Schmid2}.

However, this protection is only completely valid in the case for which the two photons are produced with perfect symmetry in their polarization states, as in Eq. \ref{eq:spdcIstate}.  Given common issues in the production of this entangled state, the photons may actually be in the state
\begin{equation}
|\psi_{0,\mathrm{asymm}}\rangle = a|HH\rangle+\sqrt{1-a^2}|VV\rangle
\label{eq:imperfectstate}
\end{equation} 
with $a^2 \neq 1/2$.  This state may be the overall two-photon state produced, or the emitted state may be symmetric but with correlations between polarization and other degrees of freedom (such as energy or spatial mode) which make it possible for an eavesdropper to filter her detection so she is dealing with an asymmetric state.  In either case, the difference in the probability of detecting the signal photon in its two polarization states can give information to an eavesdropper.  An eavesdropper, Eve, can gather this information by splitting off some photons, and using a beamsplitter to measure half of them in the $|\pm x\rangle$ basis and the other half in the $|\pm y\rangle$ basis.

\subsection{Attack on Correlation-Based Protocol}
\label{subsec:Corr-BasedAttacked}

To quantify this attack, let us assume that Eve is eavesdropping on the correlation-based protocol, just after the sender has applied a phase shift to the photon.  We want to find the probability that for a given $a^2 \neq 1/2$ and $n$ photons measured by Eve, Eve can distinguish what state the sender has prepared.  If she can reliably determine the qubit state, she has intercepted the secret.  Eve will attempt to distinguish the qubit state by counting the number of photons detected in each of her four detectors (corresponding to measured photon state $|+x\rangle$, $|-x\rangle$, $|+y\rangle$ and $|-y\rangle$).  She will then guess the bit value associated with the detector which registered the greatest number of counts.  This is not the only possible algorithm for deciding which bit value Eve will guess; however, this algorithm does yield better-than-random results for Eve whenever the vulnerability $a^2 \neq 1/2$ exists.

We assume for the sake of simplicity that the sender applies 0 phase shift, and thus Eve intercepts the signal photon which is the second member of the entangled state
\begin{equation}
|\psi_1\rangle = a|+x,H\rangle+\sqrt{1-a^2}|-x,V\rangle.
\label{eq:send0}
\end{equation}
(If the sender applies a different phase shift, the polarization states will be different, but Eve's success in identifying the correct bit does not change.)  Given that the idler photon remains unmeasured, we can then find the probability that Eve measures the signal photon in either $|+x\rangle$ or $|-x\rangle$, as well as in either $|+y\rangle$ or $|-y\rangle$.  If the photon goes to the $|\pm x\rangle$-basis detectors, the probability of registering $|+x\rangle$ is
\begin{equation}
p(|+x\rangle)=\langle\psi_1|P_{+x,s}\otimes I_i|\psi_1\rangle.
\label{eq:prob+x}
\end{equation}
This evaluates simply to $a^2$, and similarly the probability of registering $|-x\rangle$ is $1-a^2$.  If the photon goes to the $|\pm y\rangle$-basis detectors, the probability of measuring $|+y\rangle$ is
\begin{align}
p(|+y\rangle) = &\langle\psi_1|P_{+y,s}\otimes I_i|\psi_1\rangle \\
= &\frac{a^2 + (1-a^2)}{2} \notag \\
=&1/2, \notag 
\label{eq:prob+y}
\end{align}
and the probability of measuring $|-y\rangle$ is likewise $1/2$.  For the eavesdropper, then, for any single intercepted photon the probabilities of detection in $|+x\rangle$, $|-x\rangle$, $|+y\rangle$, and $|-y\rangle$ are $a^2/2$, $(1-a^2)/2$, $1/4$, and $1/4$, respectively.

Let us assume that $n$ total photons can be diverted and detected by Eve.  The probability of registering $i$ counts in the $|+x\rangle$ detector, $j$ counts in the $|-x\rangle$ detector, $k$ counts in the $|+y\rangle$ detector, and $l$ counts in the $|-y\rangle$ detector is
\begin{equation}
c_{i,j,k,l} = \frac{n!}{i!j!k!l!}\left(\frac{a^2}{2}\right)^i\left(\frac{1-a^2}{2}\right)^j\left(\frac{1}{4}\right)^k\left(\frac{1}{4}\right)^l.
\label{eq:cijkl} 
\end{equation}

Eve will successfully identify the secret if the $|+x\rangle$ detector registers the most counts ($i>j,k,l$), but also if the $|+y\rangle$ detector registers the most counts ($l>i,j,k$), since either $+$ result maps to the same secret bit value.  To exactly predict Eve's success rate, however, we must also consider the possibility that two or more detectors tie for the most counts.  In this case, we let Eve randomly select one of the tying detectors and base her bit-value guess on that detector.  With this strategy in mind, we can assign to each outcome a probability of occurring, and a likelihood for Eve to succeed in that case.

\begin{figure}[tb]
	\begin{center}
		\includegraphics[width=3.125in]{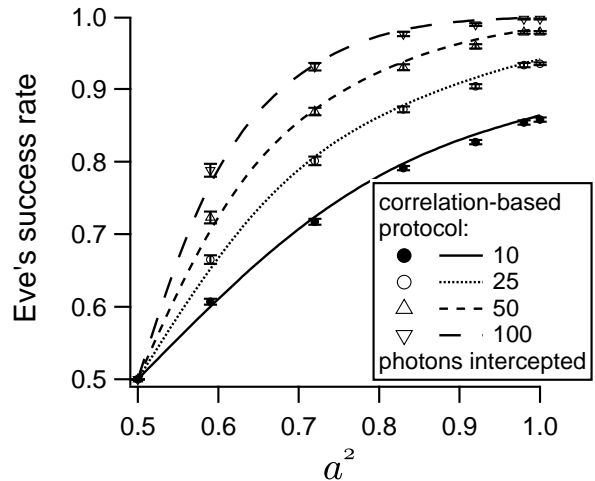}
		\caption{Bitwise probability of successful secret identification by Eve, using a photon-number splitting attack on the correlation-based SQQSS protocol.  Eve's success rate is shown as a function of $a^2$, the probability of $|HH\rangle$ in the emitted entangled state (as filtered by Eve's detection).  Because the calculation is symmetric in $a^2$ and $b^2 = 1-a^2$, $P(1-a^2)=P(a^2)$ can be used to give the probability of success for $a^2<0.5$.  Eve's success rates are shown for $n=(10,25,50,100)$ where $n$ is the number of photons detected in Eve's apparatus for each qubit.   An eavesdropper detecting an asymmetric two-photon initial state with a good signal-to-noise ratio (points) achieves results in close agreement with theory (curves).
		\label{fig:schmid_eve}}
	\end{center}
\end{figure}

Thus we can calculate the probability $p_{\mathrm{Eve},n}$ for Eve to successfully identify a bit by intercepting $n=i+j+k+l$ photons to be ($p_{i}$ indicates the probability that $i>j,k,l$, while $p_{i=j}$ indicates the probability that $i=j>k,l$, and so forth):
\begin{align}
p_{\mathrm{Eve},n} = &\frac{1}{2}\left[1+p_{i}-p_{j}+p_{i=k}-p_{j=k}\right] \notag \\
& +\frac{1}{6}\left[p_{i=k=l}-p_{j=k=l}\right] \notag \\
\approx &\frac{1}{2}\left[1+p_{i}-p_{j}\right], \text{  for large }n.
\label{eq:probderivation}
\end{align}
The final approximation simply neglects ties between the detectors, which become rare in the limit of large $n$ \cite{footnote}.  Finally, we use Eq. \ref{eq:cijkl} to evaluate the probabilities $p_{i}$, etc. in Eq. \ref{eq:probderivation}.  A plot of Eve's predicted success rates as a function of $a^2$, for various numbers of `picked-off' photons per qubit, is shown in Fig. \ref{fig:schmid_eve}.

We have focused on Eve's success rate when $a^2>1/2$; however, if $a^2<1/2$ Eve will have exactly the same success rate if she systematically reverses her bit-guessing strategy.  Thus Eve can be successful as long as (i) some $|HH\rangle$ \textit{vs.} $|VV\rangle$ asymmetry exists in the initially produced two-photon state and (ii) she is able to independently gauge the success of a small number of her guesses in order to decide whether or not to reverse her guessing strategy (a common tactic in codebreaking scenarios).  If Eve can listen in on classical communications between the secret-sharing participants, for instance, the runs used for checking error rate should allow her to determine the correct guessing strategy.

It is clear that photon number splitting requires access to many photons.  For $a^2=0.6$, measuring 100 photons gives Eve a theoretical 78\% success rate for determining the secret bit value.  The attack is especially strong when a pulse of large, indefinite photon number must be sent for each qubit.

\subsection{Robustness of Entanglement-Based Protocol}
\label{subsec:Ent-BasedNotAttacked}

The correlation-based protocol for type-I SPDC is particularly vulnerable to our PNS attack because state asymmetry of the type denoted by Eq. \ref{eq:imperfectstate} with $a^2 \neq 1/2$ is particularly common.  For example, it tends to arise from unequal thicknesses or orientations of the two crystals used for SPDC, or from imprecision in the input polarization of the SPDC pump beam.  The same type of asymmetric initial state arises in type-II SPDC, for example from imperfect alignment of the signal and idler paths with the positions of perfect overlap of the type-II output cones.  Even in systems for which the overall state has good symmetry, more troubling is the existence of correlations between photon polarization in the H/V basis and other degrees of freedom, particularly for pulsed sources \cite{KimGriceobs,Avenhaus,Poh07}, which otherwise provide an added advantage of well-defined pulse arrival times.  Correlations of this sort allow the eavesdropper to measure an effectively asymmetric state by filtering her detection on a second degree of freedom.  A typical defense against this issue is to strongly pre-filter the entangled state at its source, but this solution drastically reduces source brightness and secret-sharing transmission rate.  Schemes have been proposed to eliminate unwanted correlations without loss of brightness \cite{KimGrice02,Erdmann,Branning}, and improvements via these schemes have begun to be demonstrated experimentally \cite{Poh09,Hodelin}.  However, it is of interest to explore an approach that improves the security of single-qubit quantum secret sharing without recourse to either a loss of brightness or these additional complications.

Asymmetry between the two terms in the initial superposition or in Eve's detected state gives rise to a bias in the signal photon's polarization state -- even when observed without coincidence detection.  This bias is exploited by the PNS eavesdropper.  By contrast, if $a^2 \neq 1/2$ in the (possibly filtered) state but the entanglement-based protocol is followed, the single-particle state of the signal photon remains randomly distributed between $|+x\rangle$ and $|-x\rangle$ before the actions of the SQQSS participants.  This can be seen quite simply by rewriting the state in the $\left\{|+x\rangle,|-x\rangle\right\}$ basis for each photon:
\begin{align}	|\psi_{0,\mathrm{asymm}}\rangle = &a|HH\rangle+\sqrt{1-a^2}|VV\rangle \notag \\
= &\frac{1}{2}\left(a+\sqrt{1-a^2}\right)\left(|+x,+x\rangle+|-x,-x\rangle\right) \\ +&\frac{1}{2}\left(a-\sqrt{1-a^2}\right)\left(|+x,-x\rangle+|-x,+x\rangle\right) . \notag
	\label{eq:imperfectinx}  
\end{align}
It can easily be seen from this expression that the signal photon is found in $|+x\rangle$ 50\% of the time and $|-x\rangle$ 50\% of the time, if no polarization information is gathered for the idler.

Indeed, if the initial entangled state were asymmetric in the $x$ basis, there would be a parallel PNS eavesdropping attack, but asymmetries in the $x$ basis are much less likely simply because of the physical way in which the entangled pair is produced via SPDC.  Therefore the entanglement-based protocol, while it makes the fidelity of the transmission more sensitive, also provides built-in security against common exploitations by an eavesdropper.

\section{Experimental Implementation}
\label{Expt}

\subsection{Realization of Correlation-Based SQQSS}

\begin{figure}[t]
	\begin{center}
		\includegraphics[width=3.125in, trim = 0in 0in 0in 0in, clip=true]{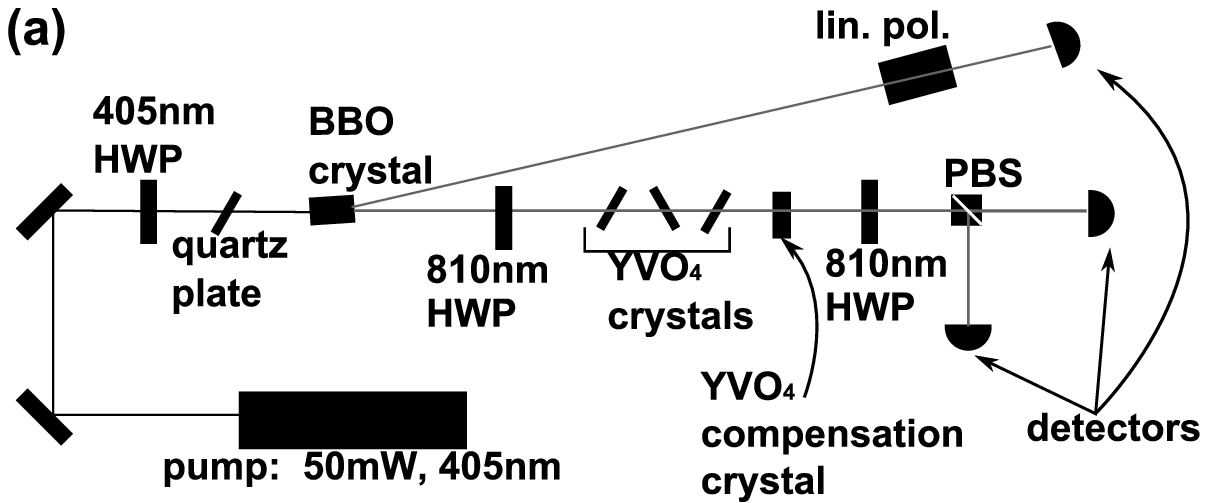}
		\includegraphics[width=3.125in, trim = 0in 0in 0in 0in, clip=true]{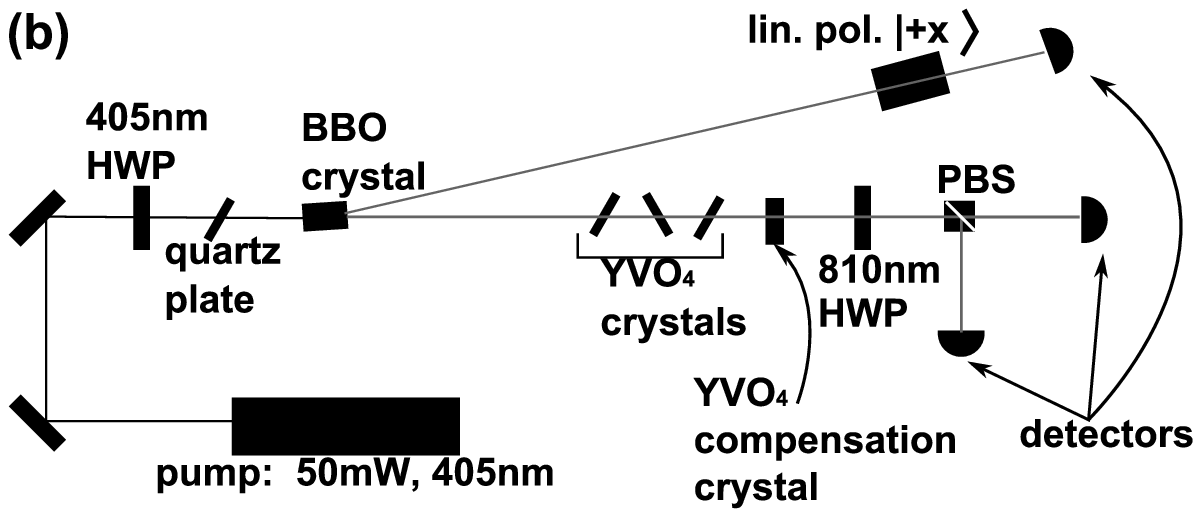}
		\caption{(a) Experimental setup for the correlation-based protocol. The sender is composed of the first 810-nm HWP along with the first YVO$_4$ crystal.  Additional YVO$_4$ crystals are the phase-shifting recipients in the protocol. The idler polarizer is set to accept horizontal polarization, which projects the signal photon into $|H\rangle$ at its entry into the SQQSS chain.  (b) In the entanglement-based protocol, the idler polarizer is set to accept $|+x\rangle$ polarization, projecting the signal photon into $|+x\rangle$ at its entry into the SQQSS chain.  The sender prepares the state using just the first YVO$_4$ crystal.
		\label{fig:mod_schmid_hmc}}
	\end{center}
\end{figure}

We now turn to implementations of both correlation-based and entanglement-based protocols with a secret sender and two recipients.  An experimental schematic for the correlation-based SQQSS is shown in Fig. \ref{fig:mod_schmid_hmc}.  Entangled photon pairs are produced via type-I degenerate spontaneous parametric downconversion \cite{typeIent} in a pair of 0.5mm-thick BBO crystals pumped with a 50mW cw diode laser at 405nm.  A 405-nm half waveplate and a tiltable quartz phase plate control the input pump beam polarization; these are set to prepare the initial entangled state of Eq. \ref{eq:spdcIstate} for the signal and idler photons at 810nm.  The BBO crystals are cut at 29.15$^{\circ}$ for noncollinear downconversion, with the signal and idler at 3$^{\circ}$ from the pump beam path.  

A Glan-Thompson polarizer in the idler beam path allows selective detection of $|H\rangle$ for the idler photon.  An 810-nm half waveplate in the signal arm converts $|H\rangle$ to $|+x\rangle$.
Now the sender and two recipients apply phase shifts $\varphi_j \in \{0,\pi/2,\pi,3\pi/2\}$.  Each phase shift is accomplished using a 200-micron-thick uniaxial YVO$_4$ crystal.  The tilt of each crystal about a vertical axis is controlled to obtain the desired phase shift.  A compensation YVO$_4$ crystal corrects for time spreading between the polarizations.  The final recipient measures the signal photon polarization in the $|\pm x\rangle$ basis using an 810nm half waveplate and a polarizing beamsplitter, sending signal photons into one of two detectors.  Signal and idler photons are detected by coupling into multimode fibers en route to single-photon counting modules and coincidence detection with a time resolution of 4ns.  The overall efficiency of detection is approximately 2\%.   

\begin{figure}[t]
	\begin{center}
		\includegraphics[width=3.125in, trim = 0in 0in 0in 0in, clip=true]{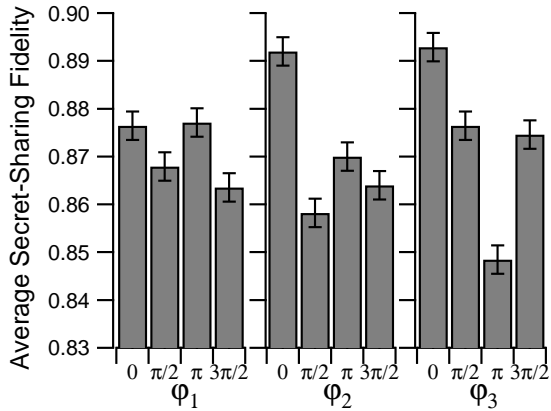}
		\caption{Experimentally observed fidelity, or probability of `sharing' the correct qubit, using the entanglement-based protocol.  Fidelity is displayed as a function of the phase-shifting angle of each YVO$_4$ phase plate, averaged over all settings of the other YVO$_4$ phase plates.  The overall 13\% qubit error rate is well below the 25\% rate for reliable detection of cheaters.  Variation of fidelity between phase plate settings matches predictions based on precision of phase plate tilting.  The overall fidelity matches predictions based on phase spread in the initial two-photon entangled state. \label{fig:successrates}}
	\end{center}
\end{figure}

\subsection{Realization of Entanglement-Based SQQSS}

An experimental schematic for the entanglement-based protocol is shown in Fig. \ref{fig:mod_schmid_hmc}.  The polarizer in the idler arm is rotated to select $|+x\rangle$ idler polarization.  The 810nm half waveplate in the idler arm is no longer necessary, so the sender is realized entirely by the first tiltable YVO$_4$ phase plate.  All other aspects of the setup remain unchanged from the correlation-based experiment.

To measure the rate of success in secret sharing, many runs with different secret and shadow bits were carried out using automated experiment control and data acquisition.  For each run, secret and shadow bits, as well as the auxiliary Class $X/Y$ bits, were chosen randomly.  The random number choices determined settings for the YVO$_4$ crystals, which were tilted using software-controlled motorized rotation platforms.  All measurements were done by counting coincidences between the idler channel and the two signal channels over a minimum time interval of 0.1 s, limited by data acquisition techniques.  Per-photon probabilities of detection were calculated, when necessary, from the observed coincidence rates. 

Observed success rates for the entanglement-based protocol are shown in Fig. \ref{fig:successrates}.  The overall error rate of 13\% is well below the 25\% threshold for reliable detection of cheaters, so secret-sharing has been realized in this implementation.  Variations in success rate of 3-6\% are observed from one set of phase plate settings to another.  This variation is predicted from the limited precision of tilt angles for the YVO$_4$ phase plate crystals. 

\begin{figure}[tb]
	\centering
		\includegraphics[width=3.125in]{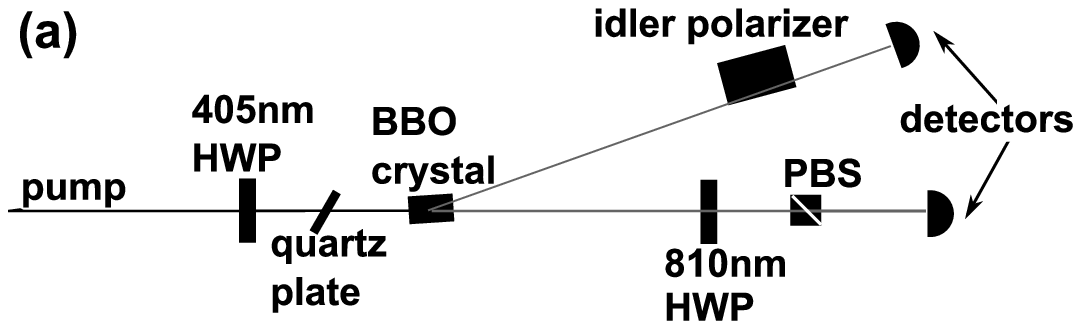}
		\includegraphics[width=3.125in]{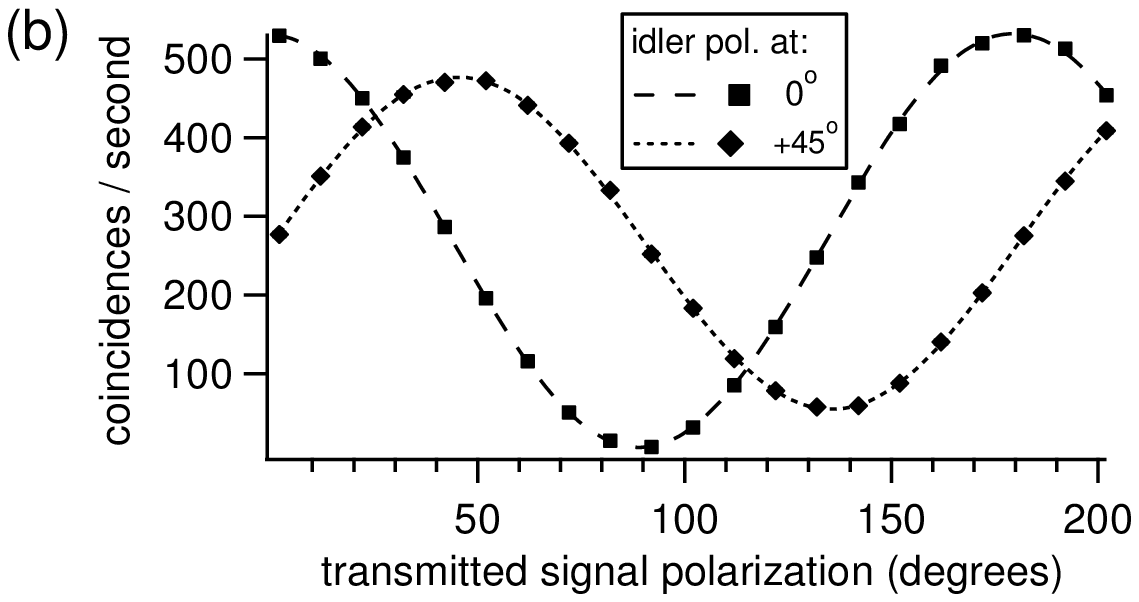}
	\caption{(a) Schematic for measuring purity (mixedness) of two-photon entangled state, due to uncompensated spread in phase $\phi$ of state $\frac{1}{\sqrt{2}}(|HH\rangle + e^{i\phi}|VV\rangle)$.  Coincidence counts between signal and idler photons are measured with the idler linear polarizer fixed at $0^\circ$ or $45^\circ$; a half waveplate in the signal arm is rotated to change the linear polarization transmitted through the polarizing beamsplitter.  (b) Coincidence counts observed.  Diminished fringe visibility with the idler polarizer at $45^\circ$ gives a purity of 0.78 for the two-photon state, accounting for our observed secret-sharing error rate.}
	\label{fig:purity}
\end{figure}

The imperfect 87\% fidelity of secret sharing can be attributed to imperfect entanglement, or purity $P<1$, of our entangled state produced by SPDC.  For our apparatus, the BBO crystal thickness and pump laser bandwidth produce a spread in phase between $|HH\rangle$ and $|VV\rangle$ components, so that our two-photon state is only partially entangled.  Separate measurements of the entangled state (see Fig. \ref{fig:purity}) indicate a purity $P=0.78$, or entangled-state fidelity $F=0.87$ \cite{fidelity}, consistent with the 87\% secret-sharing fidelity we measure in the experiment.

\subsection{Eavesdropping}

\begin{figure}[t]
	\begin{center}
		\includegraphics[width=3.125in, trim = 0in 0in 0in 0in, clip=true]{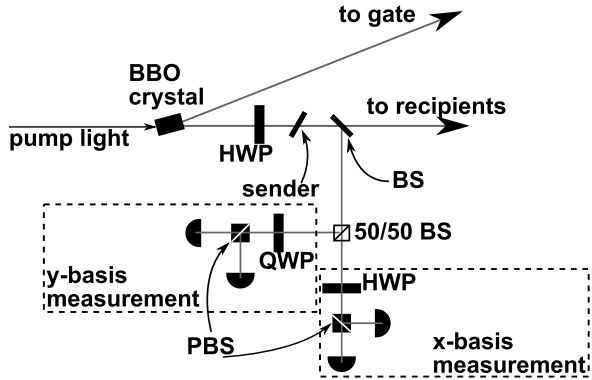}
		\caption{Ideal eavesdropping setup for photon number splitting attack.  Eve picks off a fraction of the transmitted photons and measures half of them in the $|\pm x\rangle$ basis, half of them in the $|\pm y\rangle$ basis.  If the correlation-based protocol is used, Eve can exploit an imperfectly prepared initial state to determine the secret bit with better than random success.\label{ideal_eve}}
	\end{center}
\end{figure}

A maximal implementation of the photon number splitting attack is shown in Fig. \ref{ideal_eve}.  First, some fraction of the photons traveling through the apparatus would be picked off through the use of a polarization-preserving beamsplitter placed immediately after the first (sender) YVO$_4$ phase plate. The picked-off photons would be directed into Eve's polarization-analyzing detection apparatus.  A 50-50 beam splitter (BS) sends half of the intercepted photons to be measured in the $|\pm x\rangle$ basis and the other half to be measured in the $|\pm y\rangle$ basis.  The measurement in the $|\pm x\rangle$ basis is done using a half-wave plate, polarizing beamsplitter, and two detectors. The measurement in the $|\pm y\rangle$ basis is done using a quarter-wave plate, polarizing beamsplitter, and two detectors. Eve then counts the number of photons that entered each detector and guesses a secret bit based on the detector registering the most counts.

To implement a photon number splitting attack experimentally, we simulated the success of the Fig. \ref{ideal_eve} eavesdropper via a simplified experimental setup. Rather than performing SQQSS detection as well as simultaneous detection in two bases by Eve, we omitted the final SQQSS measurement and conducted measurements in each of Eve's two bases at different times in the actual experiment. 

\begin{figure}[tb]
	\begin{center}
		\includegraphics[width=3.125in, trim = 0in 0in 0in 0in, clip=true]{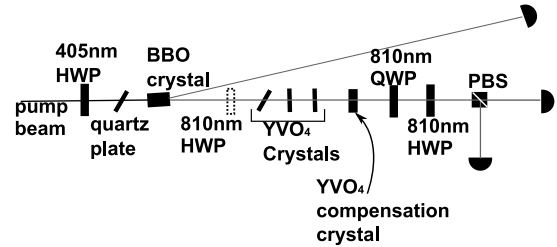}
		\caption{Experimental setup to simulate a photon number splitting eavesdropping attack on the correlation-based protocol.  The 405-nm HWP is rotated for different values of $a^2$ in the initial state.  The last two YVO$_4$ crystals are held at a constant position to give a phase shift of $\varphi=0$.  For eavesdropping on the entanglement-based protocol, the first 810-nm HWP is removed.  Detection is carried out in coincidence with the idler photon to improve signal-to-noise, but no idler polarizer is used to simulate an eavesdropper with no access to idler state information.}
		\label{hmc_eve}
	\end{center}
\end{figure}

The modified eavesdropping schematic is shown in Fig.~\ref{hmc_eve}.  The 405-nm half waveplate was rotated to obtain different values of $a^2$.  The sender's 810-nm half waveplate was present for the correlation-based protocol only.  Two of the three YVO$_4$ crystals, left in place for convenience, were held at a constant position to introduce a phase shift of 0$^{\circ}$. The first YVO$_4$ crystal was tilted to produce the four possible phase shifts introduced by the sender.  An 810-nm quarter waveplate at 45$^{\circ}$ allowed measurement in the $|\pm y\rangle$ vs. $|\pm x\rangle$ basis to be decided by rotation of the final 810-nm half waveplate.

To approximate an eavesdropper with perfect signal-to-noise on her polarization analysis, we measured the signal photon in coincidence with the idler, but with no idler polarizer since the eavesdropper lacks access to the idler's polarization information.  Thus the coincidence detection improved signal-to-noise problems due to background light, but otherwise faithfully simulated an eavesdropper's action.

A single run of the eavesdropping scheme consists of:  tilting the sender phase plate, setting the eavesdropper half waveplate to collect count rates in $|\pm x\rangle$ for 50 s, and then rotating the eavesdropper half waveplate to collect count rates in $|\pm y\rangle$ for 50 s.  The observed count rates for each sender phase setting were then used as inputs to a Monte Carlo simulation of Eve's success rate for a small number of detected photons.

\begin{figure}[tb]
	\centering
		\includegraphics[width=3.125in]{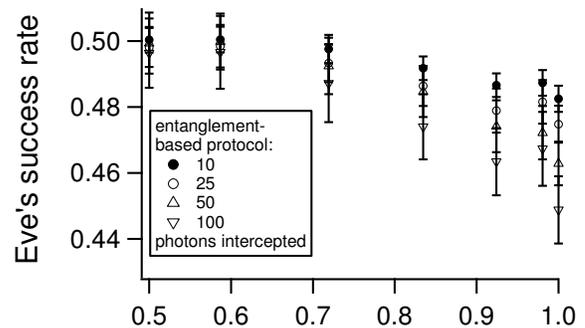}
	\caption{A plot of the success rate of an eavesdropper detecting $n$ = 10, 25, 50, or 100 photons per qubit, for the entanglement-based protocol.  Note the change of vertical scale from Fig. \ref{fig:schmid_eve}.  Here the eavesdropper is no longer successful despite asymmetry in the initial two-photon state.  The very small deviations from 50\% success can be attributed to imperfect alignment of the apparatus.}
	\label{fig:HMCEavesCoinc}
\end{figure}

Observed success rates for Eve are shown here for the correlation-based measurement (Fig. \ref{fig:schmid_eve}) and likewise for the entanglement-based measurement (Fig. \ref{fig:HMCEavesCoinc}).  The imperfect entangled-state purity seen in Fig. \ref{fig:purity} does not affect eavesdropping success.  Hence the experimentally inferred success rates closely follow the theoretical predictions discussed above. 

\section{Conclusions}

We have demonstrated two SQQSS protocols using entangled photon pairs from type-I SPDC.  A photon-number splitting attack on these protocols, exploiting lossy transmission and asymmetric state preparation, is demonstrated.  The entanglement-based scheme is robust against this attack while the correlation-based scheme is not; this contrast illustrates the value of using quantum entanglement for security of communication.

An imperfect entangled state ($P<1$) causes lowered fidelity of secret-sharing for the entanglement-based scheme.  Specifically, in this work, $P=0.78$, input state fidelity $F=0.87$, leads to secret-sharing fidelity of 87\%.  Improved purity at the $P\approx 0.95$ level should be possible in this apparatus with relatively minor modification, such as addition of a compensation BBO crystal in the pump beam to remove time/energy phase spread \cite{compensation1, compensation2, compensation3}.  Such an improvement should lead to a secret-sharing fidelity of $\approx$97\%, at which point imprecision in component positioning will limit the overall fidelity in practice.

To further probe the strengths and weaknesses of entanglement-secured SQQSS, other attack strategies must be developed and implemented, or a more general security analysis including both internal and external attacks must be conducted.  Further work can also address scaling of SQQSS with the number of participants, and variations of SQQSS in which information is encoded in multiple entangled degrees of freedom of the photon pair. 

\section{Acknowledgments}

The authors thank M. Beck at Whitman College for discussions and much valuable information regarding experimental apparatus, C. Schmid for correspondence about the correlation-based SQQSS scheme in type-II downconversion, and D.K. Koh for discussions of secret-sharing security.  This work was supported by Research Corporation Cottrell College Science Grant No. 10598, and by the Beckman Foundation through Harvey Mudd College.  P.S. acknowledges support from the Fannie and John Hertz Foundation, R.R. acknowledges support from the Engman Foundation, and D.B. acknowledges support from the R. C. Baker Foundation.  

%\bibliography{ssBib}

\begin{thebibliography}{30}
\expandafter\ifx\csname natexlab\endcsname\relax\def\natexlab#1{#1}\fi
\expandafter\ifx\csname bibnamefont\endcsname\relax
  \def\bibnamefont#1{#1}\fi
\expandafter\ifx\csname bibfnamefont\endcsname\relax
  \def\bibfnamefont#1{#1}\fi
\expandafter\ifx\csname citenamefont\endcsname\relax
  \def\citenamefont#1{#1}\fi
\expandafter\ifx\csname url\endcsname\relax
  \def\url#1{\texttt{#1}}\fi
\expandafter\ifx\csname urlprefix\endcsname\relax\def\urlprefix{URL }\fi
\providecommand{\bibinfo}[2]{#2}
\providecommand{\eprint}[2][]{\url{#2}}

\bibitem[{\citenamefont{Hillery et~al.}(1999)\citenamefont{Hillery,
  Bu\v{z}ek, and Berthiaume}}]{qss_orig}
\bibinfo{author}{\bibfnamefont{M.}~\bibnamefont{Hillery}},
  \bibinfo{author}{\bibfnamefont{V.}~\bibnamefont{Bu\v{z}ek}}, \bibnamefont{and}
  \bibinfo{author}{\bibfnamefont{A.}~\bibnamefont{Berthiaume}},
  \bibinfo{journal}{Phys. Rev. A} \textbf{\bibinfo{volume}{59}},
  \bibinfo{pages}{1829} (\bibinfo{year}{1999}).

\bibitem[{\citenamefont{Bouwmeester et~al.}(1999)\citenamefont{Bouwmeester,
  Pan, Daniell, Weinfurter, and Zeilinger}}]{multipartite1}
\bibinfo{author}{\bibfnamefont{D.}~\bibnamefont{Bouwmeester}},
  \bibinfo{author}{\bibfnamefont{J.-W.} \bibnamefont{Pan}},
  \bibinfo{author}{\bibfnamefont{M.}~\bibnamefont{Daniell}},
  \bibinfo{author}{\bibfnamefont{H.}~\bibnamefont{Weinfurter}},
  \bibnamefont{and}
  \bibinfo{author}{\bibfnamefont{A.}~\bibnamefont{Zeilinger}},
  \bibinfo{journal}{Phys. Rev. Lett.} \textbf{\bibinfo{volume}{82}},
  \bibinfo{pages}{1345} (\bibinfo{year}{1999}).

\bibitem[{\citenamefont{Eibl et~al.}(2004)\citenamefont{Eibl, Kiesel,
  Bourennane, Kurtsiefer, and Weinfurter}}]{multipartite2}
\bibinfo{author}{\bibfnamefont{M.}~\bibnamefont{Eibl}},
  \bibinfo{author}{\bibfnamefont{N.}~\bibnamefont{Kiesel}},
  \bibinfo{author}{\bibfnamefont{M.}~\bibnamefont{Bourennane}},
  \bibinfo{author}{\bibfnamefont{C.}~\bibnamefont{Kurtsiefer}},
  \bibnamefont{and}
  \bibinfo{author}{\bibfnamefont{H.}~\bibnamefont{Weinfurter}},
  \bibinfo{journal}{Phys. Rev. Lett.} \textbf{\bibinfo{volume}{92}},
  \bibinfo{pages}{077901} (\bibinfo{year}{2004}).

\bibitem[{\citenamefont{Zhao et~al.}(2004)\citenamefont{Zhao, Chen, Zhang,
  Yang, Briegel, and Pan}}]{multipartite3}
\bibinfo{author}{\bibfnamefont{Z.}~\bibnamefont{Zhao}},
  \bibinfo{author}{\bibfnamefont{Y.-A.} \bibnamefont{Chen}},
  \bibinfo{author}{\bibfnamefont{A.-N.} \bibnamefont{Zhang}},
  \bibinfo{author}{\bibfnamefont{T.}~\bibnamefont{Yang}},
  \bibinfo{author}{\bibfnamefont{H.~J.} \bibnamefont{Briegel}},
  \bibnamefont{and} \bibinfo{author}{\bibfnamefont{J.-W.} \bibnamefont{Pan}},
  \bibinfo{journal}{Nature} \textbf{\bibinfo{volume}{430}}, \bibinfo{pages}{54}
  (\bibinfo{year}{2004}).

\bibitem[{\citenamefont{Häffner et~al.}(2005)\citenamefont{H\"affner, H\"ansel,
  Roos, Benhelm, al~kar, Chwalla, K\"orber, Rapol, Riebe, Schmidt
  et~al.}}]{multipartite4}
\bibinfo{author}{\bibfnamefont{H.}~\bibnamefont{H\"affner}},
  \bibinfo{author}{\bibfnamefont{W.}~\bibnamefont{H\"ansel}},
  \bibinfo{author}{\bibfnamefont{C.~F.} \bibnamefont{Roos}},
  \bibinfo{author}{\bibfnamefont{J.}~\bibnamefont{Benhelm}},
  \bibinfo{author}{\bibfnamefont{D.~C.} \bibnamefont{al~kar}},
  \bibinfo{author}{\bibfnamefont{M.}~\bibnamefont{Chwalla}},
  \bibinfo{author}{\bibfnamefont{T.}~\bibnamefont{K\"orber}},
  \bibinfo{author}{\bibfnamefont{U.~D.} \bibnamefont{Rapol}},
  \bibinfo{author}{\bibfnamefont{M.}~\bibnamefont{Riebe}},
  \bibinfo{author}{\bibfnamefont{P.~O.} \bibnamefont{Schmidt}},
  \bibnamefont{et~al.}, \bibinfo{journal}{Nature}
  \textbf{\bibinfo{volume}{438}}, \bibinfo{pages}{643} (\bibinfo{year}{2005}).

\bibitem[{\citenamefont{Lu et~al.}(2007)\citenamefont{Lu, Zhou, G\"uhne, Gao,
  Zhang, Yuan, Goebel, Yang, and Pan}}]{multipartite5}
\bibinfo{author}{\bibfnamefont{C.-Y.} \bibnamefont{Lu}},
  \bibinfo{author}{\bibfnamefont{X.-Q.} \bibnamefont{Zhou}},
  \bibinfo{author}{\bibfnamefont{O.}~\bibnamefont{G\"uhne}},
  \bibinfo{author}{\bibfnamefont{W.-B.} \bibnamefont{Gao}},
  \bibinfo{author}{\bibfnamefont{J.}~\bibnamefont{Zhang}},
  \bibinfo{author}{\bibfnamefont{Z.-S.} \bibnamefont{Yuan}},
  \bibinfo{author}{\bibfnamefont{A.}~\bibnamefont{Goebel}},
  \bibinfo{author}{\bibfnamefont{T.}~\bibnamefont{Yang}}, \bibnamefont{and}
  \bibinfo{author}{\bibfnamefont{J.-W.} \bibnamefont{Pan}},
  \bibinfo{journal}{Nature Physics} \textbf{\bibinfo{volume}{3}},
  \bibinfo{pages}{91} (\bibinfo{year}{2007}).

\bibitem[{\citenamefont{Schmid et~al.}(2005)\citenamefont{Schmid, Trojek,
  Bourennane, Kurtsiefer, \.{Z}ukowski, and
  Weinfurter}}]{Schmid}
\bibinfo{author}{\bibfnamefont{C.}~\bibnamefont{Schmid}},
  \bibinfo{author}{\bibfnamefont{P.}~\bibnamefont{Trojek}},
  \bibinfo{author}{\bibfnamefont{M.}~\bibnamefont{Bourennane}},
  \bibinfo{author}{\bibfnamefont{C.}~\bibnamefont{Kurtsiefer}},
  \bibinfo{author}{\bibfnamefont{M.}~\bibnamefont{\.{Z}ukowski}}, \bibnamefont{and}
  \bibinfo{author}{\bibfnamefont{H.}~\bibnamefont{Weinfurter}},
  \bibinfo{journal}{Phys. Rev. Lett.} \textbf{\bibinfo{volume}{95}},
  \bibinfo{pages}{230505} (\bibinfo{year}{2005}).

\bibitem[{\citenamefont{Schmid et~al.}(2006)\citenamefont{Schmid, Trojek,
  Gaertner, Bourennane, Kurtsiefer, \.{Z}ukowski,
  and Weinfurter}}]{Schmid2}
\bibinfo{author}{\bibfnamefont{C.}~\bibnamefont{Schmid}},
  \bibinfo{author}{\bibfnamefont{P.}~\bibnamefont{Trojek}},
  \bibinfo{author}{\bibfnamefont{S.}~\bibnamefont{Gaertner}},
  \bibinfo{author}{\bibfnamefont{M.}~\bibnamefont{Bourennane}},
  \bibinfo{author}{\bibfnamefont{C.}~\bibnamefont{Kurtsiefer}},
  \bibinfo{author}{\bibfnamefont{M.}~\bibnamefont{\.{Z}ukowski}}, \bibnamefont{and}
  \bibinfo{author}{\bibfnamefont{H.}~\bibnamefont{Weinfurter}},
  \bibinfo{journal}{Fortschr. Phys.} \textbf{\bibinfo{volume}{54}},
  \bibinfo{pages}{831} (\bibinfo{year}{2006}).

\bibitem[{\citenamefont{Schmid et~al.}(2007)\citenamefont{Schmid, Trojek,
  Bourennane, Kurtsiefer, \.{Z}ukowski, and
  Weinfurter}}]{Schmid_corr}
\bibinfo{author}{\bibfnamefont{C.}~\bibnamefont{Schmid}},
  \bibinfo{author}{\bibfnamefont{P.}~\bibnamefont{Trojek}},
  \bibinfo{author}{\bibfnamefont{M.}~\bibnamefont{Bourennane}},
  \bibinfo{author}{\bibfnamefont{C.}~\bibnamefont{Kurtsiefer}},
  \bibinfo{author}{\bibfnamefont{M.}~\bibnamefont{\.{Z}ukowski}}, \bibnamefont{and}
  \bibinfo{author}{\bibfnamefont{H.}~\bibnamefont{Weinfurter}},
  \bibinfo{journal}{Phys. Rev. Lett.} \textbf{\bibinfo{volume}{98}},
  \bibinfo{pages}{028902} (\bibinfo{year}{2007}).

\bibitem[{\citenamefont{He and Wang}(2010)}]{HeWang}
\bibinfo{author}{\bibfnamefont{G.~P.} \bibnamefont{He}} \bibnamefont{and}
  \bibinfo{author}{\bibfnamefont{Z.~D.} \bibnamefont{Wang}},
  \bibinfo{journal}{Quantum Information and Computation}
  \textbf{\bibinfo{volume}{10}}, \bibinfo{pages}{28} (\bibinfo{year}{2010}).

\bibitem[{\citenamefont{Mayers}(2001)}]{qcrypt}
\bibinfo{author}{\bibfnamefont{D.}~\bibnamefont{Mayers}}, \bibinfo{journal}{J.
  ACM} \textbf{\bibinfo{volume}{48}}, \bibinfo{pages}{351}
  (\bibinfo{year}{2001}).

\bibitem[{\citenamefont{Brassard et~al.}(2000)\citenamefont{Brassard,
  L\"utkenhaus, Mor, and Sanders}}]{practical_qcrypt1}
\bibinfo{author}{\bibfnamefont{G.}~\bibnamefont{Brassard}},
  \bibinfo{author}{\bibfnamefont{N.}~\bibnamefont{L\"utkenhaus}},
  \bibinfo{author}{\bibfnamefont{T.}~\bibnamefont{Mor}}, \bibnamefont{and}
  \bibinfo{author}{\bibfnamefont{B.~C.} \bibnamefont{Sanders}},
  \bibinfo{journal}{Phys. Rev. Lett.} \textbf{\bibinfo{volume}{85}},
  \bibinfo{pages}{1330} (\bibinfo{year}{2000}).

\bibitem[{\citenamefont{Lo and L\"utkenhaus}(2007)}]{practical_qcrypt2}
\bibinfo{author}{\bibfnamefont{H.-K.} \bibnamefont{Lo}} \bibnamefont{and}
  \bibinfo{author}{\bibfnamefont{N.}~\bibnamefont{L\"utkenhaus}},
  \bibinfo{howpublished}{quant-ph/0702202} (\bibinfo{year}{2007}).

\bibitem[{\citenamefont{Gisin et~al.}(2006)\citenamefont{Gisin, Fasel, Kraus,
  Zbinden, and Ribordy}}]{trojan}
\bibinfo{author}{\bibfnamefont{N.}~\bibnamefont{Gisin}},
  \bibinfo{author}{\bibfnamefont{S.}~\bibnamefont{Fasel}},
  \bibinfo{author}{\bibfnamefont{B.}~\bibnamefont{Kraus}},
  \bibinfo{author}{\bibfnamefont{H.}~\bibnamefont{Zbinden}}, \bibnamefont{and}
  \bibinfo{author}{\bibfnamefont{G.}~\bibnamefont{Ribordy}},
  \bibinfo{journal}{Phys. Rev. A} \textbf{\bibinfo{volume}{73}},
  \bibinfo{pages}{022320} (\bibinfo{year}{2006}).

\bibitem[{\citenamefont{Rosenberg et~al.}(2007)\citenamefont{Rosenberg,
  Harrington, Rice, Hiskett, Peterson, Hughes, Lita, Nam, and
  Nordholt}}]{decoys1}
\bibinfo{author}{\bibfnamefont{D.}~\bibnamefont{Rosenberg}},
  \bibinfo{author}{\bibfnamefont{J.~W.} \bibnamefont{Harrington}},
  \bibinfo{author}{\bibfnamefont{P.~R.} \bibnamefont{Rice}},
  \bibinfo{author}{\bibfnamefont{P.~A.} \bibnamefont{Hiskett}},
  \bibinfo{author}{\bibfnamefont{C.~G.} \bibnamefont{Peterson}},
  \bibinfo{author}{\bibfnamefont{R.~J.} \bibnamefont{Hughes}},
  \bibinfo{author}{\bibfnamefont{A.~E.} \bibnamefont{Lita}},
  \bibinfo{author}{\bibfnamefont{S.~W.} \bibnamefont{Nam}}, \bibnamefont{and}
  \bibinfo{author}{\bibfnamefont{J.~E.} \bibnamefont{Nordholt}},
  \bibinfo{journal}{Phys. Rev. Lett.} \textbf{\bibinfo{volume}{98}},
  \bibinfo{pages}{010503} (\bibinfo{year}{2007}).

\bibitem[{\citenamefont{Schmitt-Manderbach
  et~al.}(2007)\citenamefont{Schmitt-Manderbach, Weier, F\"urst, Ursin,
  Tiefenbacher, Scheidl, Perdigues, Sodnik, Kurtsiefer, Rarity
  et~al.}}]{decoys2}
\bibinfo{author}{\bibfnamefont{T.}~\bibnamefont{Schmitt-Manderbach}},
  \bibinfo{author}{\bibfnamefont{H.}~\bibnamefont{Weier}},
  \bibinfo{author}{\bibfnamefont{M.}~\bibnamefont{F\"urst}},
  \bibinfo{author}{\bibfnamefont{R.}~\bibnamefont{Ursin}},
  \bibinfo{author}{\bibfnamefont{F.}~\bibnamefont{Tiefenbacher}},
  \bibinfo{author}{\bibfnamefont{T.}~\bibnamefont{Scheidl}},
  \bibinfo{author}{\bibfnamefont{J.}~\bibnamefont{Perdigues}},
  \bibinfo{author}{\bibfnamefont{Z.}~\bibnamefont{Sodnik}},
  \bibinfo{author}{\bibfnamefont{C.}~\bibnamefont{Kurtsiefer}},
  \bibinfo{author}{\bibfnamefont{J.~G.} \bibnamefont{Rarity}},
  \bibnamefont{et~al.}, \bibinfo{journal}{Phys. Rev. Lett.}
  \textbf{\bibinfo{volume}{98}}, \bibinfo{pages}{010504}
  (\bibinfo{year}{2007}).

\bibitem[{foo()}]{footnote}
\bibinfo{note}{Furthermore, this approximation always leads to a bias against
  Eve's true chances of success, because ties between $|+x\rangle$ and
  $|+y\rangle$ (which always result in the correct qubit) are more common than
  ties between $|-x\rangle$ and $|-y\rangle$ (which result in the incorrect
  qubit) as long as $a^2 > 1/2$.}

\bibitem[{\citenamefont{Kim and Grice}(2005)}]{KimGriceobs}
\bibinfo{author}{\bibfnamefont{Y.}~\bibnamefont{Kim}} \bibnamefont{and}
  \bibinfo{author}{\bibfnamefont{W.}~\bibnamefont{Grice}},
  \bibinfo{journal}{Opt. Lett.} \textbf{\bibinfo{volume}{30}},
  \bibinfo{pages}{908} (\bibinfo{year}{2005}).

\bibitem[{\citenamefont{Avenhaus et~al.}(2009)\citenamefont{Avenhaus, Chekhova,
  Krivitsky, Leuchs, and Silberhorn}}]{Avenhaus}
\bibinfo{author}{\bibfnamefont{M.}~\bibnamefont{Avenhaus}},
  \bibinfo{author}{\bibfnamefont{M.~V.} \bibnamefont{Chekhova}},
  \bibinfo{author}{\bibfnamefont{L.~A.} \bibnamefont{Krivitsky}},
  \bibinfo{author}{\bibfnamefont{G.}~\bibnamefont{Leuchs}}, \bibnamefont{and}
  \bibinfo{author}{\bibfnamefont{C.}~\bibnamefont{Silberhorn}},
  \bibinfo{journal}{Phys. Rev. A} \textbf{\bibinfo{volume}{79}},
  \bibinfo{pages}{043836} (\bibinfo{year}{2009}).

\bibitem[{\citenamefont{Poh et~al.}(2007)\citenamefont{Poh, Lum, Marcikic,
  Lamas-Linares, and Kurtsiefer}}]{Poh07}
\bibinfo{author}{\bibfnamefont{H.~S.} \bibnamefont{Poh}},
  \bibinfo{author}{\bibfnamefont{C.~Y.} \bibnamefont{Lum}},
  \bibinfo{author}{\bibfnamefont{I.}~\bibnamefont{Marcikic}},
  \bibinfo{author}{\bibfnamefont{A.}~\bibnamefont{Lamas-Linares}},
  \bibnamefont{and}
  \bibinfo{author}{\bibfnamefont{C.}~\bibnamefont{Kurtsiefer}},
  \bibinfo{journal}{Phys. Rev. A} \textbf{\bibinfo{volume}{75}},
  \bibinfo{pages}{043816} (\bibinfo{year}{2007}).

\bibitem[{\citenamefont{Kim and Grice}(2002)}]{KimGrice02}
\bibinfo{author}{\bibfnamefont{Y.}~\bibnamefont{Kim}} \bibnamefont{and}
  \bibinfo{author}{\bibfnamefont{W.}~\bibnamefont{Grice}}, \bibinfo{journal}{J.
  Mod. Opt.} \textbf{\bibinfo{volume}{49}}, \bibinfo{pages}{2309}
  (\bibinfo{year}{2002}).

\bibitem[{\citenamefont{Erdmann et~al.}(2000)\citenamefont{Erdmann, Branning,
  Grice, and Walmsley}}]{Erdmann}
\bibinfo{author}{\bibfnamefont{R.}~\bibnamefont{Erdmann}},
  \bibinfo{author}{\bibfnamefont{D.}~\bibnamefont{Branning}},
  \bibinfo{author}{\bibfnamefont{W.}~\bibnamefont{Grice}}, \bibnamefont{and}
  \bibinfo{author}{\bibfnamefont{I.~A.} \bibnamefont{Walmsley}},
  \bibinfo{journal}{Phys. Rev. A} \textbf{\bibinfo{volume}{62}},
  \bibinfo{pages}{053810} (\bibinfo{year}{2000}).

\bibitem[{\citenamefont{Branning et~al.}(1999)\citenamefont{Branning, Grice,
  Erdmann, and Walmsley}}]{Branning}
\bibinfo{author}{\bibfnamefont{D.}~\bibnamefont{Branning}},
  \bibinfo{author}{\bibfnamefont{W.~P.} \bibnamefont{Grice}},
  \bibinfo{author}{\bibfnamefont{R.}~\bibnamefont{Erdmann}}, \bibnamefont{and}
  \bibinfo{author}{\bibfnamefont{I.~A.} \bibnamefont{Walmsley}},
  \bibinfo{journal}{Phys. Rev. Lett.} \textbf{\bibinfo{volume}{83}},
  \bibinfo{pages}{955} (\bibinfo{year}{1999}).

\bibitem[{\citenamefont{Poh et~al.}(2009)\citenamefont{Poh, Lim, Marcikic,
  Lamas-Linares, and Kurtsiefer}}]{Poh09}
\bibinfo{author}{\bibfnamefont{H.~S.} \bibnamefont{Poh}},
  \bibinfo{author}{\bibfnamefont{J.}~\bibnamefont{Lim}},
  \bibinfo{author}{\bibfnamefont{I.}~\bibnamefont{Marcikic}},
  \bibinfo{author}{\bibfnamefont{A.}~\bibnamefont{Lamas-Linares}},
  \bibnamefont{and}
  \bibinfo{author}{\bibfnamefont{C.}~\bibnamefont{Kurtsiefer}},
  \bibinfo{journal}{Phys. Rev. A} \textbf{\bibinfo{volume}{80}},
  \bibinfo{pages}{043815} (\bibinfo{year}{2009}).

\bibitem[{\citenamefont{Hodelin et~al.}(2006)\citenamefont{Hodelin, Khoury, and
  Bouwmeester}}]{Hodelin}
\bibinfo{author}{\bibfnamefont{J.~F.} \bibnamefont{Hodelin}},
  \bibinfo{author}{\bibfnamefont{G.}~\bibnamefont{Khoury}}, \bibnamefont{and}
  \bibinfo{author}{\bibfnamefont{D.}~\bibnamefont{Bouwmeester}},
  \bibinfo{journal}{Phys. Rev. A} \textbf{\bibinfo{volume}{74}},
  \bibinfo{pages}{013802} (\bibinfo{year}{2006}).

\bibitem[{\citenamefont{Kwiat et~al.}(1999)\citenamefont{Kwiat, Waks, White,
  Appelbaum, and Eberhard}}]{typeIent}
\bibinfo{author}{\bibfnamefont{P.~G.} \bibnamefont{Kwiat}},
  \bibinfo{author}{\bibfnamefont{E.}~\bibnamefont{Waks}},
  \bibinfo{author}{\bibfnamefont{A.~G.} \bibnamefont{White}},
  \bibinfo{author}{\bibfnamefont{I.}~\bibnamefont{Appelbaum}},
  \bibnamefont{and} \bibinfo{author}{\bibfnamefont{P.~H.}
  \bibnamefont{Eberhard}}, \bibinfo{journal}{Phys. Rev. A}
  \textbf{\bibinfo{volume}{60}}, \bibinfo{pages}{R773} (\bibinfo{year}{1999}).

\bibitem[{fid()}]{fidelity}
\bibinfo{note}{The fidelity of the experimental two-photon density matrix
  $\rho$ with respect to the ideal two-photon pure state $|\Phi^+\rangle$ is
  defined as $Tr(\sqrt{\sqrt{\rho}|\Phi^+\rangle\langle\Phi^+|\sqrt{\rho}})$,
  and is analogous to the state overlap for the case of two pure states.}

\bibitem[{\citenamefont{Nambu et~al.}(2002)\citenamefont{Nambu, Usami, Tsuda,
  Matsumoto, and Nakamura}}]{compensation1}
\bibinfo{author}{\bibfnamefont{Y.}~\bibnamefont{Nambu}},
  \bibinfo{author}{\bibfnamefont{K.}~\bibnamefont{Usami}},
  \bibinfo{author}{\bibfnamefont{Y.}~\bibnamefont{Tsuda}},
  \bibinfo{author}{\bibfnamefont{K.}~\bibnamefont{Matsumoto}},
  \bibnamefont{and} \bibinfo{author}{\bibfnamefont{K.}~\bibnamefont{Nakamura}},
  \bibinfo{journal}{Phys. Rev. A} \textbf{\bibinfo{volume}{66}},
  \bibinfo{pages}{033816} (\bibinfo{year}{2002}).

\bibitem[{\citenamefont{Altepeter et~al.}(2005)\citenamefont{Altepeter,
  Jeffrey, and Kwiat}}]{compensation2}
\bibinfo{author}{\bibfnamefont{J.~B.} \bibnamefont{Altepeter}},
  \bibinfo{author}{\bibfnamefont{E.~R.} \bibnamefont{Jeffrey}},
  \bibnamefont{and} \bibinfo{author}{\bibfnamefont{P.}~\bibnamefont{Kwiat}},
  \bibinfo{journal}{Optics Express} \textbf{\bibinfo{volume}{13}},
  \bibinfo{pages}{8951} (\bibinfo{year}{2005}).

\bibitem[{\citenamefont{Rangarajan et~al.}(2009)\citenamefont{Rangarajan,
  Goggin, and Kwiat}}]{compensation3}
\bibinfo{author}{\bibfnamefont{R.}~\bibnamefont{Rangarajan}},
  \bibinfo{author}{\bibfnamefont{M.}~\bibnamefont{Goggin}}, \bibnamefont{and}
  \bibinfo{author}{\bibfnamefont{P.}~\bibnamefont{Kwiat}},
  \bibinfo{journal}{Optics Express} \textbf{\bibinfo{volume}{17}},
  \bibinfo{pages}{18920} (\bibinfo{year}{2009}).

\end{thebibliography}

\bibliographystyle{apsrev}

\end{document}